\documentclass[a4paper, 12pt]{article}
\usepackage[dvips]{graphicx}

\def\ltap{\ \raise.3ex\hbox{$<$\kern-.75em\lower1ex\hbox{$\sim$}}\ }
\def\gtap{\ \raise.3ex\hbox{$>$\kern-.75em\lower1ex\hbox{$\sim$}}\ }

\begin{document}

\title{Chiral symmetry breaking and stability of quark droplets}
\author{S. Yasui and A. Hosaka \\
 Research Center for Nuclear Physics, Osaka University \\
 567-0047 10-1, Mihogaoka, Ibaraki, Osaka, Japan }
\date{}
\maketitle

\begin{abstract}
We discuss the stability of strangelets --- quark droplets with strangeness --- in the Nambu--Jona-Lasinio model supplemented by a boundary condition for quark confinement.
Effects of dynamical chiral symmetry breaking are considered properly inside quark droplets of arbitrary baryon number.
We obtain the energy per baryon number of quark droplets with baryon number from one to thousands.
It is shown that strangelets are not the ground states as compared with nuclei, though they can be locally stable.
\end{abstract}

\section{Introduction}

The strange matter, which is a quark matter composed of the $u$, $d$ and $s$ quarks, has been one of the interesting subjects in the quark and hadron physics \cite{Bodmer, Witten, Farhi}.
This subject is also related with other fields, such as the topics of compact stars, the early universe,  heavy ion collision phenomena and so on.

However, the physics of the strange matter has been a difficult problem of QCD.
Effective models of QCD were used from the early stage.
The MIT bag model has been often used to show the stability of the strange matter \cite{Bodmer, Witten, Farhi, Madsen}.
In this model, it is assumed that the strange matter is approximated as almost free quark gas with current quark masses.
There, it was shown that the strange matter can be stable as compared with the $ud$ quark matter.
This is because the $s$ quark current mass is small as compared with the Fermi energy of the $ud$ quark.
Moreover, it was argued that the strange matter could be more stable than nuclear matter.
Quark droplets with finite volume with strangeness, which we call strangelets, could be also more stable than atomic nuclei.
If the strangelets are stable particles, they should be observed as almost neutral heavy particles in experiments.
Indeed, several candidates of exotic particles have been observed in the cosmic rays \cite{Banerjee, Kasuya, Ichimura, Price, Miyamura, Capdevielle, Weber}.
There are also experiments to search for such exotic particles in heavy ion collision experiments \cite{Weber, Rischke, Bielich}.

In the arguments in \cite{Bodmer, Witten, Farhi, Madsen}, it was assumed that chiral symmetry is restored in the strange matter.
However, when we allow chiral symmetry breaking, the stability of the strange matter is not necessarily concluded.
This is because $s$ quarks acquire a large dynamical mass by chiral symmetry breaking.
In \cite{Buballa96, Buballa98, Buballa99}, the stability of the bulk strange matter was discussed by considering dynamical breaking of chiral symmetry in the Nambu--Jona-Lasinio (NJL) model, in which the quark dynamical mass depended on the baryon number density.
In these discussions, however, it was shown that the strange matter could not be absolutely stable.
The main reason of the instability of the strange matter is that the dynamical mass of the $s$ quarks is larger than the $ud$ quark Fermi energy.
It was pointed out that the dynamics of chiral symmetry breaking is related with the formation of quark droplets \cite{Alford}.
The stability of strangelets were discussed in a constituent quark model with one-gluon-exchange interaction \cite{Tamagaki1, Tamagaki2, Tamagaki3}.
There, it was assumed that the quarks have constant constituent mass.
Thus, chiral symmetry breaking is an important ingredient to discuss the stability of the strange matter.

It was also pointed out that chiral symmetry breaking plays an important role, not only in the bulk quark matter, but also in strangelets of finite volume \cite{Kim, Kiriyama_Hosaka, Yasui}.
In \cite{Kiriyama_Hosaka, Yasui}, it was argued that the chiral symmetry breaking in the strangelets depended on the size of the strangelet by using the NJL model supplemented by the quark confinement.
In this argument, it was assumed that the quarks interacting by the NJL interaction are confined in a MIT bag as a strangelet.
This model is called the {\it NJL+MIT bag} model.
In our previous work \cite{Kiriyama_Hosaka, Yasui}, instead of solving the boundary condition of the MIT bag exactly, the Multiple Reflection Expansion (MRE) \cite{MRE} was used as an approximate method with an expectation that the strangelets have large baryon number.
There, the surface effects were included by calculating the density of states of quarks  modified by the boundary condition.
It was shown that the chiral symmetry of the $s$ quark sector in the strangelet is restored by the surface effect up to the baryon number of thousands, and strangelets could be more stable than the $ud$ quark droplets.
However, it was not possible to compare the stability of the strangelets with baryons (nucleons, hyperons and so on), since the mass of a single baryon was not calculated due to the limitation of the MRE approximation.

In this paper, we investigate the NJL+ MIT bag model as one of the effective models of the baryon and the quark droplet.
We consider directly the discrete quark energy levels by solving the boundary condition, which implements the quark confinement.
The advantage of the present method is that we can construct a solution for all baryon number $A$ including the nucleon and hyperon.
This is contrasted with the previous discussion by the MRE which can be applied to the case of relatively large $A$ \cite{Yasui}.
In the present study, we discuss the absolute stability of the strangelets as compared with baryons.
In addition, we check the validity of the approximation of the MRE. 

Contents of this paper are the followings.
In Section 2, we formulate the NJL + MIT bag model to describe chiral symmetry breaking in a strangelet.
In Section 3, we discuss our numerical results by solving the discrete quark energy levels.
There, the efficiency of the MRE approximation is investigated. 
We discuss the masses of a single baryon and quark droplet, and compare the stability between them.
In Section 4, we conclude our discussion.

\section{Formulation}

The dynamical breaking of chiral symmetry is one of the important aspects of the low energy QCD.
As an effective theory of QCD, we consider a four-quark interaction of the NJL model for chiral symmetry breaking \cite{Nambu}.
For a quark droplet of finite size, valence quarks are assumed to exist inside the quark droplet. 
In order to implement this condition, we assume that quarks are confined in the MIT bag.
Namely, we impose the MIT bag boundary condition at the bag surface of the quark droplet.
With these assumptions, we use the lagrangian
\begin{eqnarray}
{\cal L} \!=\! \left[ \bar{\psi}(i\partial\hspace{-0.2cm}/ \!-\! m^{0}) \psi
                       \!+\! \frac{G}{2} \sum_{a\!=\!0}^{8}
					                       \left\{    \left( \bar{\psi} \lambda^{a}\psi \right)^{2}
											   \!+\! \left( \bar{\psi} i\gamma_{5} \lambda^{a}\psi \right)^{2}
										   \right\}
						\right] \theta(R-r)
                      \!-\! \frac{1}{2}\bar{\psi}\psi \delta(r\!-\!R),
\label{eq : NJL_bag}
\end{eqnarray}
where $\psi \!=\! (u, d, s)^{t}$ is the quark field, and $m^{0} \!=\! \mbox{diag}(m_{u}^{0},m_{d}^{0},m_{s}^{0})$ current quark mass matrix.
The second term in the first bracket is the NJL point-like interaction term invariant under $U(N_{f})_{L} \times U(N_{f})_{R}$, in which $\lambda^{a}$ ($a\!=\!0, \cdots, 8$) are the Gell-Mann matrices normalized by $\mbox{tr}\lambda^{a}\lambda^{b} \!=\! 2\delta^{ab}$.
We do not consider the 't~Hooft term for explicit $U(1)_{A}$ breaking.
In Eq.~(\ref{eq : NJL_bag}), the step function is multiplied in order to have quarks only inside the bag, and the last term realizes the MIT bag boundary condition for vanishing vector current flow at the bag surface \cite{Bogolioubov, Thomas, Hosaka, Hosaka_Toki}.
Formally, our model violates chiral symmetry at the bag surface.
However, it is expected to be a good approximation  that we do not consider the pion and kaon cloud for relatively large quark droplets. 
Details will be discussed in the later sections.
Here, $r$ is a distance from the center of the bag.
It is assumed that the strangelet has a spherical shape with a bag radius $R$.
We call Eq.~(\ref{eq : NJL_bag}) as the {\it NJL+MIT bag} model.
In our model setting, the outside of the bag is described by the usual NJL model, which gives the chirally broken phase as vacuum.

Chiral symmetry breaking is caused by the non-perturbative dynamics of the NJL interaction, in which $\bar{q}q$ correlation takes a non-zero expectation value.
To investigate the $\bar{q}q$ condensate inside the bag, we use a mean field approximation for the scalar channel by replacing $(\bar{q}q)^{2} \!\rightarrow\! 2\bar{q}q\langle\bar{q}q\rangle \!-\! \langle \bar{q}q \rangle^{2}$ for $q \!=\!u, d$ and $s$, respectively.
Here $\langle\bar{q}q\rangle$ is the expectation value of $\bar{q}q$. 
Then, the lagrangian is approximated as 
\begin{eqnarray}
{\cal L} &\!=\!& 
		    \sum_{q\!=\!u,d,s}
		                      \left[ 
							     \left\{
								   \bar{q}(i\partial\hspace{-0.2cm}/ \!-\! m_{q}) q
                       \!-\! \frac{\phi_{q}^{2}}{4G}
					             \right\} \theta(R-r)
			           -  \frac{1}{2}\bar{q}q \delta(r\!-\!R)
							  \right],
\label{eq : NJL_bag2}
\end{eqnarray}
where we have defined
\begin{eqnarray}
      \phi_{q} \!=\! -2G \langle \bar{q}q \rangle, 
\label{eq : constituent_mass}
\end{eqnarray}
as an order parameter for chiral symmetry breaking, which also represents the constituent quark mass.
We also have defined the dynamical mass of a quark by
\begin{eqnarray}
     m_{q} \!=\! m_{q}^{0} + \phi_{q}.
 \label{eq : dynamical_mass}
\end{eqnarray}
In this approximation, we obtain a single particle equation inside the bag ($r<R$)
\begin{eqnarray}
     (i\partial\hspace{-0.2cm}/ \!-\! m_{q}) q = 0,
      \label{eq : single_particle}
\end{eqnarray}
with the boundary condition at the bag surface ($r=R$)
\begin{eqnarray}
  -i  \vec{n}\cdot\vec{\gamma}  q = q,
  \label{eq : bc1}
\end{eqnarray}
which is obtained from the last term  in  Eq.~(\ref{eq : NJL_bag2}).
Here, $\vec{\gamma}$ is spatial part of the gamma matrices.

We solve Eq.~(\ref{eq : single_particle}) by using spherical basis set.
The quark eigenstates of momentum $p\!=\!|\vec{p}|$, total angular momentum $j\!=\!l\!+\!\epsilon/2$ $(\epsilon\!=\!\pm1)$ and parity $P\!=\!(-1)^{l}$ are given by,
\begin{eqnarray}
    \psi_{j=l\!+\!\epsilon/2 \, m}^{l}(\vec{x})
\!=\! {\cal N}
\left(
\begin{array}{c}
  j_{l}(pr)  \\
 \epsilon\frac{ip}{E+m_{q}} j_{l\!+\!\epsilon}(pr) \vec{\sigma} \!\cdot\! \vec{n}
\end{array}
\right)
{\cal Y}_{jm}^{l}(\vec{n}),
\label{eq : eigen_function}
\end{eqnarray}
where $\vec{n}$ is a unit vector with angular direction $\vec{n}\!=\!\vec{x}/r$ and $r\!=\!|\vec{x}|$.
In Eq.~(\ref{eq : eigen_function}), $j_{l}$ is the spherical Bessel function, and $\cal N$ is a normalization constant.
The spinor harmonics ${\cal Y}_{jm}^{l}(\vec{n})$, which are eigenstates of the total angular momentum $j\!=\!l\!+\!\epsilon/2$ $(\epsilon\!=\!\pm1)$, is given by 
\begin{eqnarray}
         {\cal Y}_{jm}^{l}(\vec{n}) \!=\! \sum_{\mu}( l\mu\frac{1}{2} m\!-\!\mu | j m ) Y_{l\mu}(\vec{n})\chi_{m\!-\!\mu},
\end{eqnarray}
where $ Y_{l\mu}(\vec{n})$ are spherical harmonics and $\chi_{\mu}$ two component spinors.
We write a quark energy $E \!=\! \sqrt{p^{2}\!+\!m_{q}^{2}}$ with quark mass $m_{q}$.

The boundary condition Eq.~(\ref{eq : bc1}) of this quark wave function gives the momentum $p$ in positive energy state by
\begin{eqnarray}
         j_{l}(pR) \!=\! \epsilon \frac{p}{E\!+\!m_{q}}j_{l\!+\!\epsilon}(pR),
\label{eq : boundary_condition}
\end{eqnarray}
and the momentum $p$ in negative energy state by
\begin{eqnarray}
         j_{l\!+\!\epsilon}(pR) \!=\! -\epsilon \frac{p}{E\!+\!m_{q}} j_{l}(pR).
\label{eq : boundary_condition2}
\end{eqnarray}
The absolute energy value of the state $j^{P}$ in the vacuum is as same as that of the state $j^{-P}$ in the valence. 
By solving the boundary condition (\ref{eq : boundary_condition}) and (\ref{eq : boundary_condition2}), we obtain a sequence of states of a momenta $p=p_{j^{P},n}(R, m_{q})$ with node quantum numbers $n$ as a function of the bag radius $R$ and the quark mass $m_{q}$.
The momentum $p_{j^{P},n}(R, m_{q})$ must be self-consistent with the mean field approximation Eqs.~(\ref{eq :  constituent_mass}) and (\ref{eq :  dynamical_mass}).

We obtain the energy density $\varepsilon$ of strangelets with radius $R$ by taking a sum of the energy over all states including positive and negative energy states in the bag,
\begin{eqnarray}
       \varepsilon(\phi)
&=& \sum_{q=u,d,s}
     \left[ 
                         \frac{\phi_{q}^{2}}{4G} 
              \!+\! \frac{\nu}{\frac{4}{3}\pi R^{3}} \left( 
        				        \sum_{j^{P}, n}^{p_{Fq}} E_{j^{P},n}(\phi_{q})
							 + \sum_{j^{P},n} \overline{E}_{j^{P},n}(\phi_{q}) g(\bar{p}_{j^{P},n}/\Lambda)
                          \right)
    \right]
- \varepsilon_{0},
\label{eq : energy_density_cavity}
\end{eqnarray}
where $\nu\!=\!N_{spin}\!\times\!N_{c}\!=\!2\!\times\!3\!=\!6$ is the number of degeneracy, and for positive energy state $j^{P}$,
\begin{eqnarray}
      E_{j^{P},n}(\phi_{q}) = \sqrt{ p_{j^{P},n}^{2} + (m_{q}^{0} + \phi_{q})^{2} },
\end{eqnarray}
and for negative energy state $j^{P}$,
\begin{eqnarray}
      \overline{E}_{j^{P},n}(\phi_{q}) = - \sqrt{ \bar{p}_{j^{P},n}^{2} + (m_{q}^{0} + \phi_{q})^{2} }.
\end{eqnarray}
Here, $p_{j^{P},n}$ and $\bar{p}_{j^{P},n}$ are the momenta in the valence and vacuum parts, which are given by the boundary condition (\ref{eq : boundary_condition}) and (\ref{eq : boundary_condition2}), respectively.
The Fermi momenta $p_{Fq}$ are obtained by
\begin{eqnarray}
      \sum_{j^{P}, n}^{p_{Fq}} = N_{q},
\end{eqnarray}
for a given quark number $N_{q}$ for each flavor $q\!=\!u, d$ and $s$.
Since the NJL model is a cutoff theory, we have introduced a regularization function $g(p/\Lambda)$ for a sum over negative energy states.
The results may depend on the choice of $g(p/\Lambda)$.
In the present work, we use a smooth function of the Lorentzian type 
\begin{eqnarray}
       g(p/\Lambda) &=& \frac{1}{1+(p/\Lambda)^{a}},
	   \label{eq : regularization}
\end{eqnarray}
with a diffuseness parameter $a$.
We consider that the outside of the bag is chirally broken phase descried by the NJL model.
In Eq.~(\ref{eq : energy_density_cavity}), the energy density of the strangelet $\varepsilon$ is measured from the vacuum of the chirally broken phase in the absence of the bag.
In the NJL model, the latter energy density is calculated by
\begin{eqnarray}
    \varepsilon_{0} \equiv \varepsilon_{bulk}^{vac}(\phi_{0})
    	  = \sum_{q=u,d,s}
               \left[ 
                           \frac{\phi_{q0}^{2}}{4G}
                         - \nu \int \frac{d^{3}p}{(2\pi)^{3}} g(p/\Lambda)
                                                     \sqrt{p^{2} + (m_{q}^{0} + \phi_{q0})^{2}}
               \right],
\label{eq : vacuum_bulk}
\end{eqnarray}
where $\phi_{0}=(\phi_{u0},\phi_{d0},\phi_{s0})^{t}$ is the value of the order parameter in the chirally broken vacuum.

For later discussions, we write the energy density of the strangelet as a sum of the kinetic energy part of the valence quarks and the rest,
\begin{eqnarray}
     \varepsilon(\phi) = \sum_{q=u,d,s} \sum_{j^{P}, n}^{p_{Fq}} E_{j^{P},n}(\phi_{q}) + B(\phi).
\end{eqnarray}
Here, the effective bag constant $B(\phi)$ is the energy density of the vacuum inside the bag measured from $\varepsilon_{0}$,
\begin{eqnarray}
      B(\phi) &=& \varepsilon_{bag}^{vac}(\phi) - \varepsilon_{0},
	 \label{eq : bag_constant}
\end{eqnarray}
where we define the vacuum energy inside the bag
\begin{eqnarray}
     \varepsilon_{bag}^{vac}(\phi) =  \sum_{q=u,d,s}
     \left[ 
                         \frac{\phi_{q}^{2}}{4G} 
              \!+\! \nu \sum_{j^{P},n} \overline{E}_{j^{P},n}(\phi_{q}) g(\bar{p}_{j^{P},n}/\Lambda)
    \right].
	\label{eq : vacuum_bag}
\end{eqnarray}
We mention that the effective bag constant $B$ is not a constant independent of the size of the bag, but is a quantity depending on the mass of the quarks $m_{q}$, or $\phi_{q}$, and the bag radius $R$.

By using the energy density (\ref{eq : energy_density_cavity}), the total energy of a strangelet is given by
\begin{eqnarray}
      E\!=\!\varepsilon V - \alpha/R,
\end{eqnarray}
with the volume $V\!=\!\frac{4\pi}{3}R^{3}$.
The second term of this equation expresses the correction from zero point energy, center-of-mass motion and other effects  \cite{DeGrand, Bhaduri}. 
Then, we obtain a locally stable state of a quark droplet for a given baryon number $A$ and strangeness $S$ ($S=-N_{s}$ and $N_{u}=N_{d}$).
There, the order parameter $\phi_{q}$ and the bag radius $R$ are determined by stationary condition of the energy $E$ with respect to $\phi_{q}$ ($\phi_{u}=\phi_{d}$) and $R$, respectively;
\begin{eqnarray}
       \frac{\partial E}{\partial \phi_{q}} = 0 \mbox{ and } \frac{\partial E}{\partial R}=0.
\label{eq : SD}
\end{eqnarray}

%parameters
We have determined the model parameters as follows; the three dimensional momentum cut-off $\Lambda\!=\!0.60$ GeV, the coupling constant $G\Lambda^{2}\!=\!4.61$, the current quark masses $m_{u}^{0}\!=\!m_{d}^{0}\!=\!0.0059$ GeV and $m_{s}^{0}\!=\!0.134$ GeV.
This parameter set reproduces the pion decay constant $f_{\pi}\!=\!0.093$ GeV, pion mass $m_{\pi}\!=\!0.139$ GeV, kaon mass $m_{K}=0.493$ GeV, and dynamical quark mass $m_{u}\!=\!m_{d}\!=\!0.378$ GeV in the chirally broken vacuum.
The diffuseness parameter $a$ in the regularization (\ref{eq : regularization}) is treated as a free parameter by fitting the baryon mass.
We use the value $a=22$ in Subsections 3.2 and 3.3.
In Subsection 3.3, we discuss the $a$ dependence of the baryon mass.

\section{Numerical results}

In this section, numerical results are investigated for the strangelets with the baryon number $A$.

\subsection{Comparison with MRE}

First, we compare the present result with the previous ones obtained by the Multiple Reflection Expansion (MRE).
In the MRE, the surface term in the quark droplet is introduced as a correction to the bulk quark matter \cite{MRE}.
The MRE was often used for analysis of quark droplets in the MIT bag model \cite{Madsen}.

In our previous work in \cite{Yasui}, the MRE was applied to the strangelets with baryon number $A \gtap 10$.
Now, we can check the efficiency of the MRE approximation, by comparing the number of valence quarks in the MRE with that obtained by calculating the discrete levels.
In the bag model, we solve the boundary condition  (\ref{eq : boundary_condition}) and (\ref{eq : boundary_condition2}).
Then, for a bag with a radius $R$, the total fermion number in the momentum space up to the momentum $\Lambda$ is obtained as
\begin{eqnarray}
  N_{bag}(\Lambda) = \sum_{j^{P},n} \theta(\Lambda - p_{j^{P},n}).
  \label{eq : number_discrete}
\end{eqnarray}
On the other hand, the MRE gives
\begin{eqnarray}
  N_{MRE}(\Lambda) = 2 \int_{0}^{\Lambda} \frac{p^{2}dp}{2\pi^{2}} \rho_{MRE}(p,m,R),
  \label{eq : number_MRE}
\end{eqnarray}
where the factor $2$ is the number of degeneracy of spin and $\rho_{MRE}(p,m,R)$ is the density of state of the MRE \cite{Madsen, MRE}, which is given as
\begin{eqnarray}
      \rho_{MRE}(p,m,R) = 1+  \frac{6\pi^{2}}{pR} f_{S} (p/m) +
                                         \frac{12\pi^2}{(pR)^2} f_{C}(p/m) + \cdots,
 \label{eq : MRE}
\end{eqnarray}
with
\begin{eqnarray}
           f_{S}(x) &=& -\frac{1}{8\pi} \left(1-\frac{2}{\pi} \arctan x \right),
\\ \nonumber
          f_{C}(x) &=& \frac{1}{12\pi^2}
                          \left[1-\frac{3x}{2}\left(\frac{\pi}{2}-\arctan x \right)\right].
\end{eqnarray}

\begin{figure}[tbp]
\begin{minipage}{8cm}
\vspace*{0.0cm}
\mbox{ \small {\bf (a)$R=1$} fm}
\centering
\includegraphics[width=8cm]{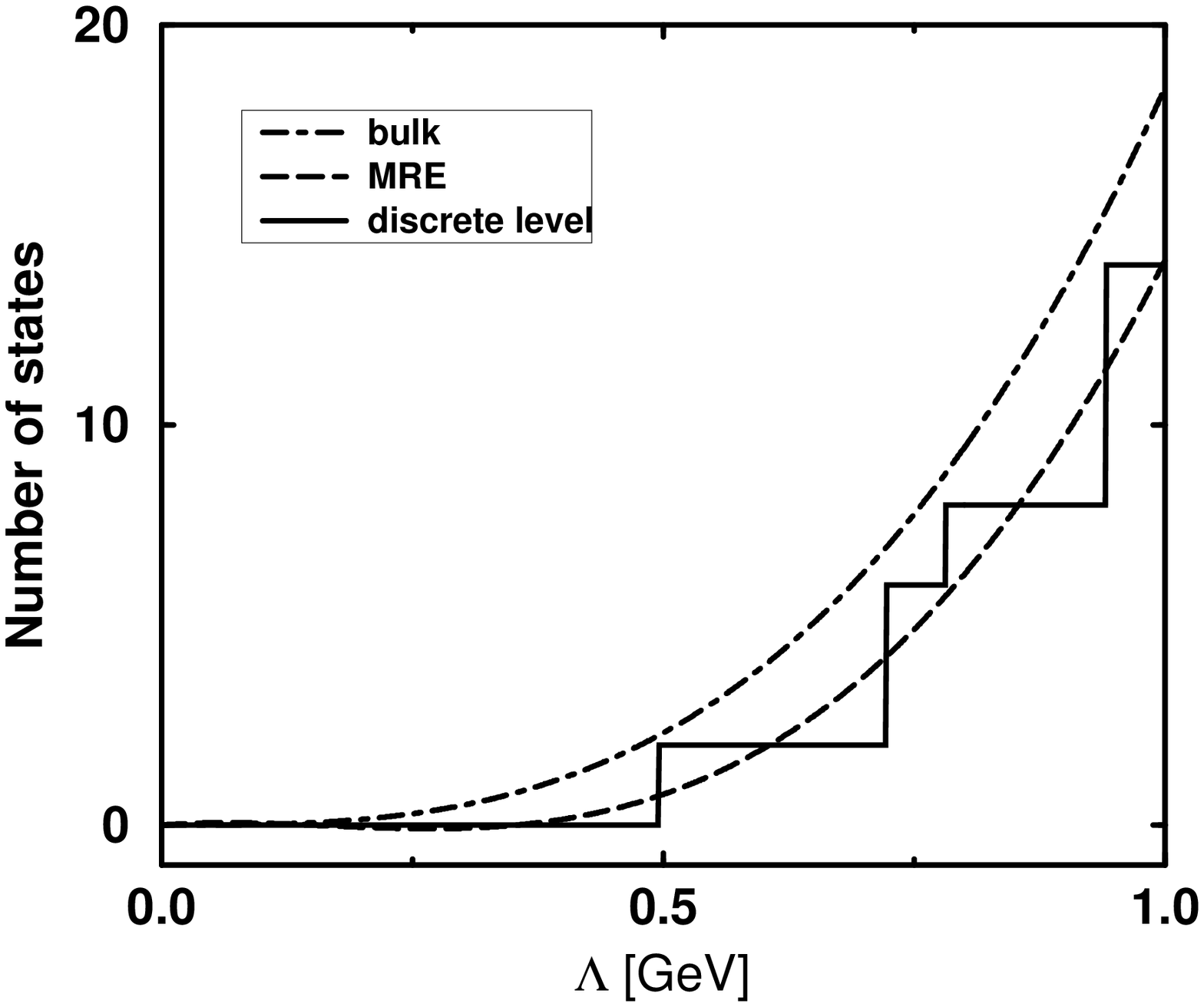}
\vspace{-0.0cm}
\end{minipage}
%%%%%%%%%%%%%%%%%%%%%%%%%%%%%%%%%
\begin{minipage}{8cm}
\mbox{ \small {\bf (b)$R=2$} fm}
\centering
\includegraphics[width=8cm]{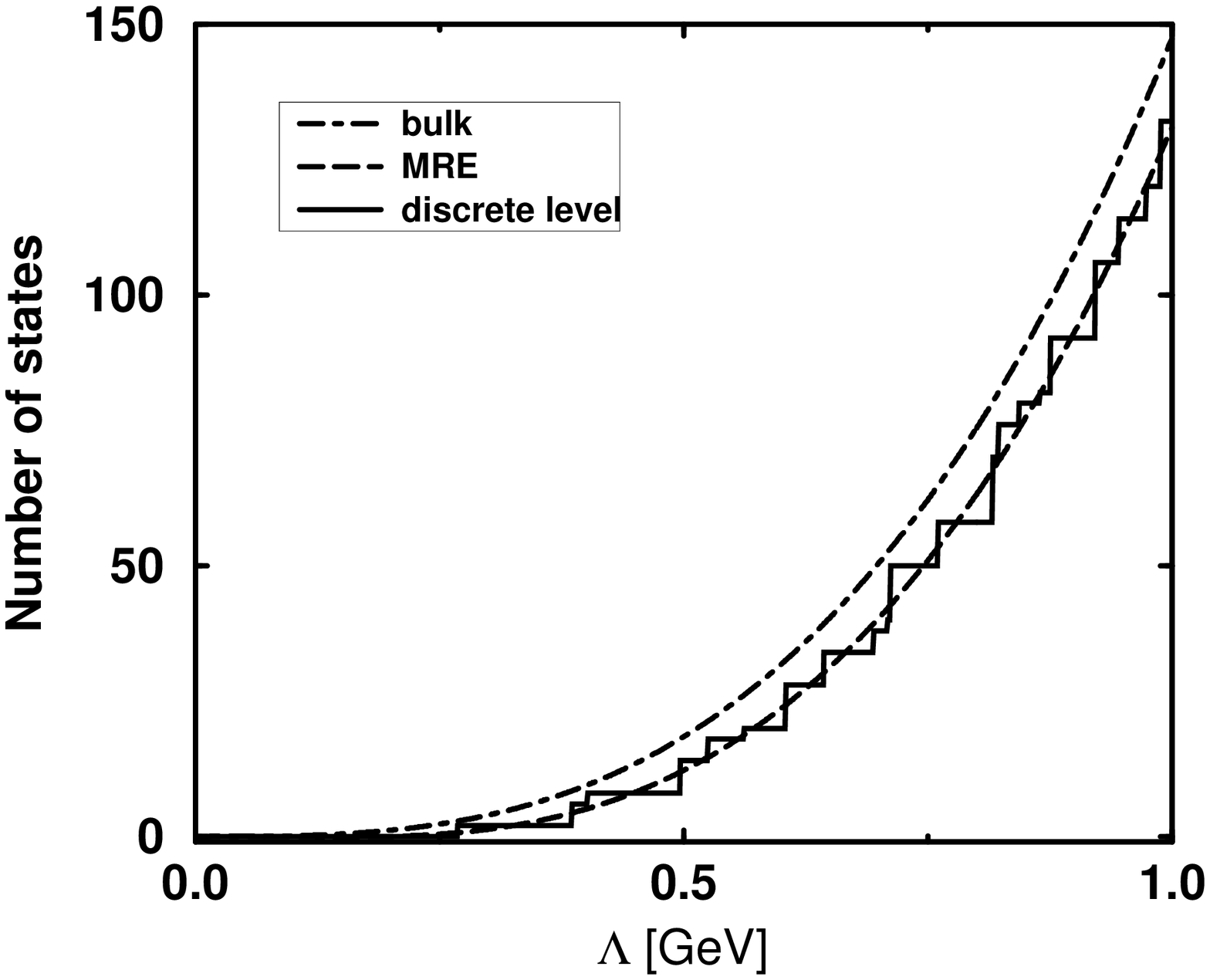}
\vspace{-0cm}
\end{minipage}
\caption{\small \baselineskip=0.5cm The number of states of quarks ($N_{bag}$ (solid lines),  $N_{MRE}$ (dashed lines) and $N_{bulk}$ (dashed-dotted lines)) in the bag as functions of  the three dimensional momentum cut-off $\Lambda$ for radii  (a) $R=1$ fm  and (b) $R=2$ fm , respectively. The fermion mass is set to be $m_{q}=0.3$ GeV.}
    \label{fig : states_p_R1_verB}
\end{figure}

We show the number of states $N_{bag}(\Lambda)$ and $N_{MRE}(\Lambda)$ for the fermion mass $m\!=\!0.3$ GeV for the bag radii $R=1$ and $2$ fm, in Fig.~\ref{fig : states_p_R1_verB}(a) and (b), respectively.
For comparison, we show also the number of states in the bulk matter 
\begin{eqnarray}
  N_{bulk}(\Lambda) = 2 \int_{0}^{\Lambda} \frac{p^{2}dp}{2\pi^{2}}.
  \label{eq : number_bulk}
\end{eqnarray}
The behavior like step function is due to the discreteness of the quark levels. 
In both figures (a) and (b), we see that the result of the MRE agrees well with the averaged behavior of the present discrete levels.
Furthermore, it is clear that the MRE approximation agrees better, as the bag radius increases.
Thus, we can use the MRE approximation instead of the discrete level calculation for larger bag radii.
In later discussions, we will confirm that the energy per baryon number of strangelets calculated from the discrete levels is consistent with that in the MRE approximation for large bag radii $R \gtap 2$ fm (see Figs.~\ref{fig : B_R_A} and \ref{fig : E_A}).

%%%%%%%%%%%%%%%%%%%%%%%%%%%%%%%%%
%%%%%%%%%%%%%%%%%%%%%%%%%%%%%%%%%%%%%%%%%%%%%%%%%%

\subsection{Vacuum structure in the bag}

The boundary conditions (\ref{eq : boundary_condition}) and (\ref{eq : boundary_condition2}) modify the quark energy levels in the bag, which is known to be an origin of the Casimir effect.
This affects not only the valence quarks with positive energy, but also the vacuum quarks with negative energy.
Accordingly, the properties of the dynamical chiral symmetry breaking is also changed.
In this subsection, we discuss the vacuum structure and the dynamics of chiral symmetry breaking in the bag . 

\begin{figure}[tbp]
\begin{minipage}{8cm}
\vspace*{0.0cm}
\mbox{ \small \bf (a)}
\centering
\includegraphics[width=8cm]{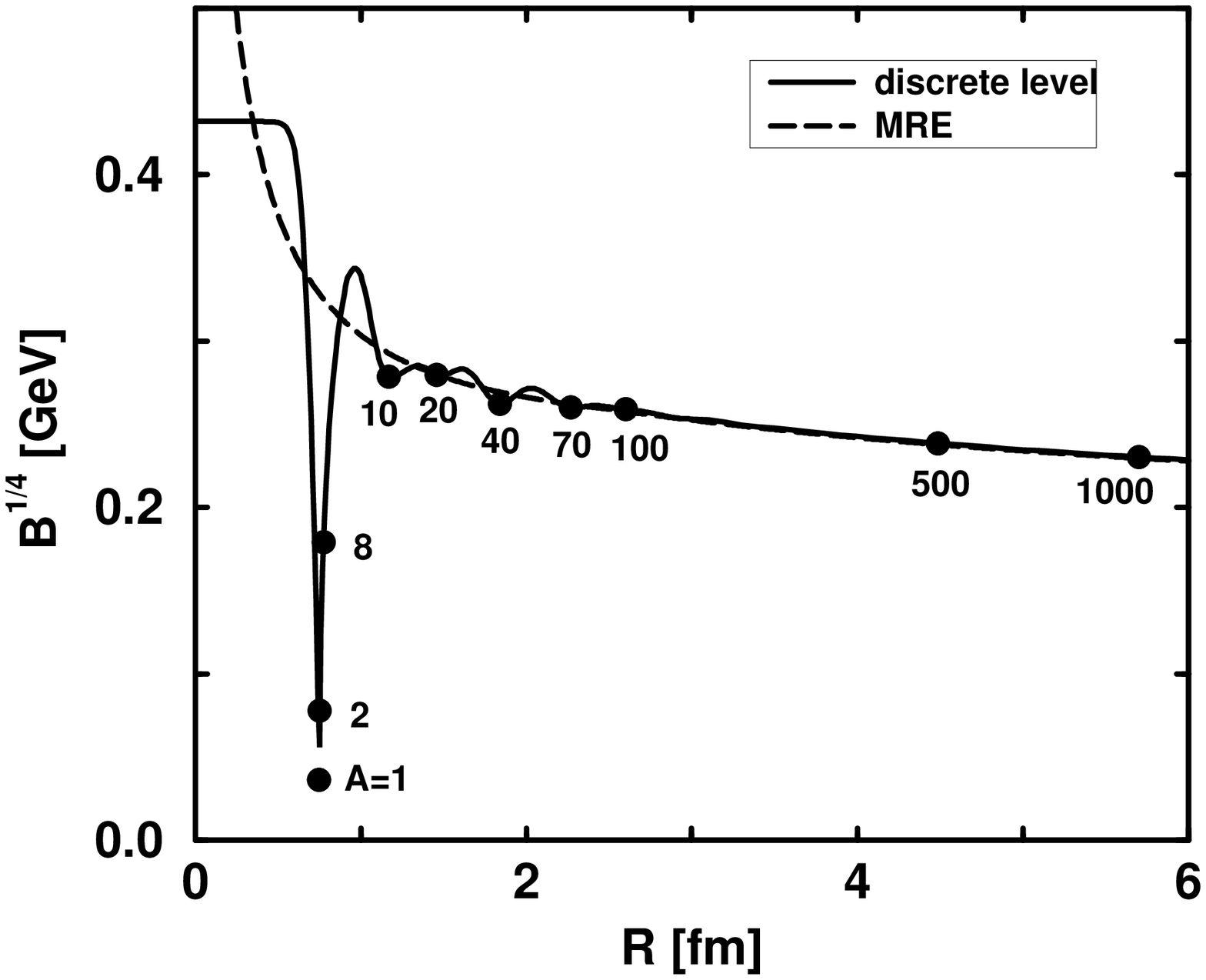}
\vspace{-0.0cm}
\end{minipage}
%%%%%%%%%%%%%%%%%%%%%%%%%%%%%%%%%
\begin{minipage}{8cm}
\mbox{ \small \bf (b)}
\centering
\includegraphics[width=8cm]{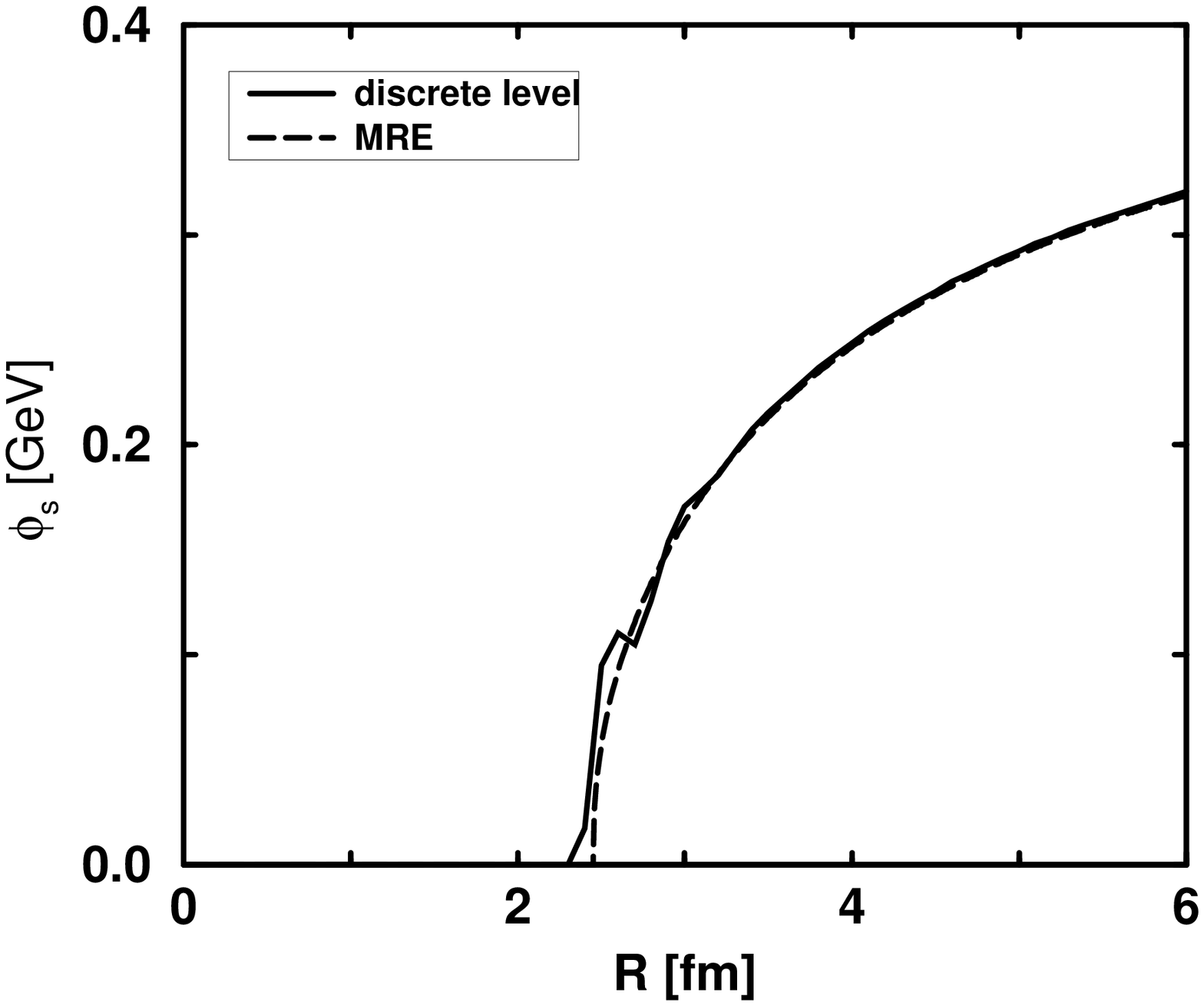}
\vspace{-0cm}
\end{minipage}
\caption{\small \baselineskip=0.5cm (a) The effective bag constant $B$ as a function of a radius of the vacuum bag.  The blobs correspond to points of baryon number $A$ as denoted by the labels. (b) The order parameter of chiral symmetry breaking in $s$ quark sector $\phi_{s}$ as a function of the bag radius $R$. In both figures, the solid lines are the present results of discrete levels, and the dashed lines are the results by the MRE method.}
    \label{fig : B_R_A}
\end{figure}

To start with, we consider an empty bag with only vacuum quarks.
The energy of the vacuum inside the bag is given by the effective bag constant $B(\phi)$ of (\ref{eq : bag_constant}). 
This quantity is obtained by taking a variation with respect to the order parameter $\phi=(\phi_{u},\phi_{d},\phi_{s})$ for the bag with a given bag radius $R$.
The resulting effective bag constant $B$ and the order parameter of chiral symmetry breaking in $s$ quark sector $\phi_{s}$ are shown in Fig.~\ref{fig : B_R_A} (a) and (b), respectively, as functions of the bag radius $R$.
The order parameter in the $ud$ sector is zero for the region of the bag radii $R \ltap 6$ fm.
We compare the results obtained by calculating the discrete levels (solid lines) and by the MRE (dashed lines).
For later convenience, we also show the effective bag constants and the radii of quark droplets filled with valence quarks by blobs.
The corresponding baryon numbers are shown in the same figure.
These values are obtained by taking a variation of the energy with respect to the quark mass and the bag radius.
The quark droplets with valence quarks are discussed in Section 3.3.

In Fig.~\ref{fig : B_R_A}(a), it is shown that the effective bag constant is a function of the bag radius $R$.
As a bulk property, the effective bag constant becomes small as the bag radius increases.
Indeed, we see a good agreement between the MRE and the present calculation with the discrete levels for larger bag radii $R \gtap 2$ fm.
The state in the bag in the limit of the large bag radius $R\rightarrow \infty$ is chirally broken bulk vacuum, because the surface effect of the bag can be neglected.
In this limit, we obtain $B=0$ form Eq.~(\ref{eq : bag_constant}) by substituting $\phi=\phi_{0}$.

On the other hand, in the limit of small bag radius $R \rightarrow 0$, the effective bag constant becomes a positive quantity.
Let us investigate slightly more of the vacuum structure in the small bag radius.
The highest state in the negative energy sea in the bag is the $1/2^{-}$ state with energy $\overline{E}_{1/2^{-}}=-2.04/R$.
As the cut-off is introduced in the bag vacuum part in Eq.~(\ref{eq :  vacuum_bag}), no level exists for  bag radii $R \ltap 2.04/\Lambda (= 0.67 \; \mbox{fm})$, since all the states are below the cut-off $\Lambda$.
Therefore, we find $\varepsilon_{bag}^{vac}(\phi) = \sum_{\phi=u,d,s}\phi^{2}/4G$ from Eq.~(\ref{eq : vacuum_bag}), which follows $\phi_{q}=0$ as chirally restored phase.
Consequently, we obtain 
\begin{eqnarray}
  B = - \varepsilon_{0}
  \hspace{1cm}
  (R \ltap 0.67 \; \mbox{fm}),
  \label{eq : function2}
\end{eqnarray}
which is a constant value independent of the bag radius.
We show that $-\varepsilon_{0}$ is a positive quantity.
For this purpose, we rewrite the vacuum energy $\varepsilon_{0}$ as 
\begin{eqnarray}
       \varepsilon_{0} = - B_{bulk}
									  + \varepsilon_{bulk}^{vac}(\phi=0),
       \label{eq : equation1}
\end{eqnarray}
where we define 
\begin{eqnarray}
    B_{bulk} = \varepsilon_{bulk}^{vac}(\phi=0) - \varepsilon_{0}.
	\label{eq : equation2}
\end{eqnarray}
This is the energy density difference between the chirally broken and restored phases in the bulk space without considering the boundary condition.
We also obtain the energy density of the vacuum quark in the chirally restored phase,
\begin{eqnarray}
   \varepsilon_{bulk}^{vac}(\phi=0) = - \nu \int \frac{d^{3}p}{(2\pi)^{2}} \sqrt{p^{2}+m_{f,0}^{2}} g(p/\Lambda).
\end{eqnarray} 
It is clear that $ \varepsilon_{bulk}^{vac}(\phi=0) $ is a negative quantity, and $B_{bulk}$ is also a positive quantity.
Therefore, $\varepsilon_{0}$ is a negative quantity.
Finally, we conclude that the effective bag constant $B=-\varepsilon_{0}$ is a positive quantity.

The above analysis seems to suggest that the effective bag constant $B$ is a monotonically decreasing function of the bag radius $R$.
However, there is a sharp decrease of the effective bag constant around the bag radius $R \simeq 0.7$ fm in the discrete level calculation, which is not shown in the MRE, as shown in Fig.~\ref{fig : B_R_A}(a).
This behavior is understood by considering the $1/2^{-}$ state in the vacuum part in the bag.
As shown in Eq.~(\ref{eq : function2}),  we have a constant effective bag constant $B=-\varepsilon_{0}$ for the bag radii $R \ltap 2.04/\Lambda$, because no quark state exists anymore in the negative energy of $|\bar{p}_{j^{P},n}|>\Lambda$.
On the other hand, for the bag radius $R \gtap 2.04/\Lambda$, several states, such as $1/2^{-}$, $3/2^{+}$ and so on,  are taken into account in the calculation of the effective bag constant.
Especially, for the bag radius $2.04/\Lambda \ltap R \ltap 3.20/\Lambda$ ($0.67\mbox{fm} \ltap R \ltap 1.05 \mbox{fm}$), only the $1/2^{-}$ state appears in Eq.~(\ref{eq : vacuum_bag}).
The next state $2/3^{+}$ with energy $\overline{E}_{3/2^{+}}=-3.20/R$ does not exist in the vacuum in this bag.
Therefore, for these bag radii, we obtain the effective bag constant 
\begin{eqnarray}
   B \simeq - \frac{ 2.04N_{1/2^{-}}}{RV} g(2.04/\Lambda R) - \epsilon_{0} 
   \hspace{1cm}
   (0.67 \; \mbox{fm} \ltap R \ltap 1.05 \; \mbox{fm}),
  \label{eq : function3}
\end{eqnarray}
where $N_{1/2^{-}} = 18$ is the number of degeneracy of $1/2^{-}$ state including the $u$, $d$ and $s$ quarks, $V=\frac{4\pi}{3}R^{3}$ the volume of the bag.
Here, chiral symmetry of all flavors is restored as shown in Fig.~\ref{fig : B_R_A}(b), and quark masses are current ones.
The energy of $u$, $d$ and $s$ quarks are almost degenerate, since the masses are smaller than the quark momenta inside the bag, $p \sim 2/R \sim 0.5 $ GeV $>>m_{u,d,s}^{0}$.
Now, the effective bag constant (\ref{eq : function3}) is an increasing function $B \sim -1/R^{4}$ of the bag radius $R$.
Consequently, from Eqs.~(\ref{eq : function2}) and (\ref{eq : function3}), we obtain the sharp decrease of the effective bag constant $B$ around $R\simeq 0.7$ fm.

%%%%%%%%%%%%%%%%%%%%%%%%%%%%%%%%%%%%%%%%%%%%%%%
%%%%%%%%%%%%%%%%%%%%%%%%%%%%%%%%%%%%%%%%%%%%%%%%%%

\subsection{Stability of strangelets}

In this subsection, we discuss the stability of strangelets with baryon number $A$ and strangeness $S$.
First, we show a global behavior of the quark droplets including the baryon.
Next, we discuss the structure of quark droplets in detail.

\subsubsection{General properties of quark droplets}

\begin{figure}[ptb]
\begin{center}
\includegraphics[width=8cm, height=8cm, keepaspectratio, clip]{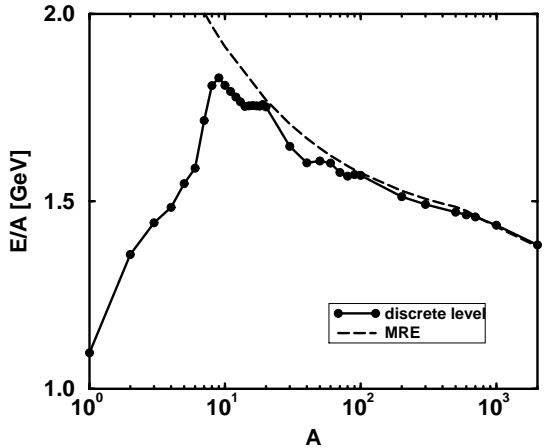}
\end{center}
\vspace*{-1.0cm} \caption{\small \baselineskip=0.5cm The energy per baryon number $E/A$ as a function of baryon number $A$. The solid line is by the discrete energy level, and the dashed line is by the MRE method.}
 \label{fig : E_A}
\end{figure}

First, we outline our result of the baryon number dependence of the properties of the quark droplets.
In Fig.~\ref{fig : E_A}, we show the resulting energy per baryon number $E/A$ as a function of baryon number $A$, which is obtained by taking a variation with respect to the chiral condensate $\phi_{u}$, $\phi_{d}$ and $\phi_{s}$, the bag radius $R$.
We also consider the variation with respect to the strangeness $S=-N_{s}$.
The solid line is obtained by considering the discrete energy levels, and the dashed line is the previous result obtained by the MRE method \cite{Yasui}.
Here, we define the strangeness fraction $r_{s}$ as a ratio of the $s$ quark number to the total quark number, $r_{s}=N_{s}/3A$.

From Fig.~\ref{fig : E_A}, in the large baryon number $A\gtap 10$, the present result of the discrete levels agrees well with that in the MRE.
This justifies the validity of using the MRE method for the baryon number $A \gtap 10$ \cite{Yasui}.
On the other hand, in the small baryon number $A\ltap 10$, the MRE does not agree with the present calculation.
The discrete level calculation reproduces nucleon mass $m_{N}=1.10$ GeV at $A=1$.
Instead, the mass of the quark droplet in the MRE becomes an unphysical value above 2 GeV as the baryon number approaches $A=1$.

\begin{figure}[ptb]
\begin{center}
\includegraphics[width=8cm, height=8cm, keepaspectratio, clip]{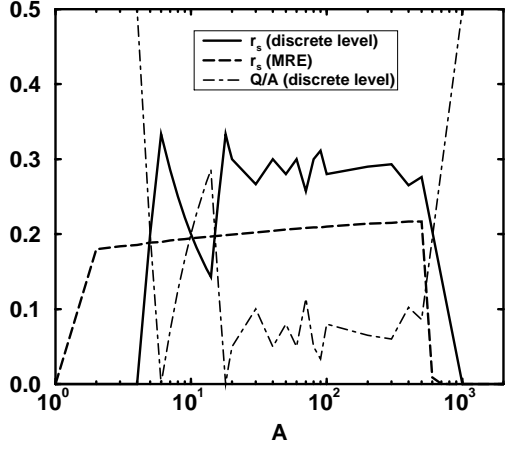}
\end{center}
\vspace*{-1.0cm} \caption{\small \baselineskip=0.5cm The strangeness fraction $r_{s}$ as a function of the baryon number $A$. The solid line is by the discrete energy level, and the long-dashed line is by the MRE method. The dot-dashed line shows the electric charge per baryon number $Q/A$ in the discrete level calculation.}
 \label{fig : rs_A}
\end{figure}

We show the resulting strangeness fraction $r_{s}$ as a function of the baryon number $A$ in Fig.~\ref{fig : rs_A}.
Here, we define the strangeness fraction $r_{s}$ as a ratio of the $s$ quark number to the total quark number, $r_{s}=N_{s}/3A$.
It is shown that the quark droplets with baryon number $5 \leq A \ltap 10^{3}$ have the finite strangeness fraction $r_{s} \sim 0.3$.
Therefore, the strangelets can exist as local minima for the limited baryon number $5 \leq A \ltap 10^{3}$.
On the other hand, the ground state of the quark droplets with baryon number $A \gtap 10^{3}$ is the $ud$ quark matter.
The value of the critical baryon number $A \simeq 10^{3}$  was also obtained in the calculation by the MRE \cite{Yasui}.
The quark droplets in the limit of the large baryon number is the bulk quark matter of $ud$ quarks.
This is consistent with the results obtained by the NJL model \cite{Buballa96, Buballa98, Buballa99}.

\begin{figure}[ptb]
\begin{center}
\includegraphics[width=8cm, height=8cm, keepaspectratio, clip]{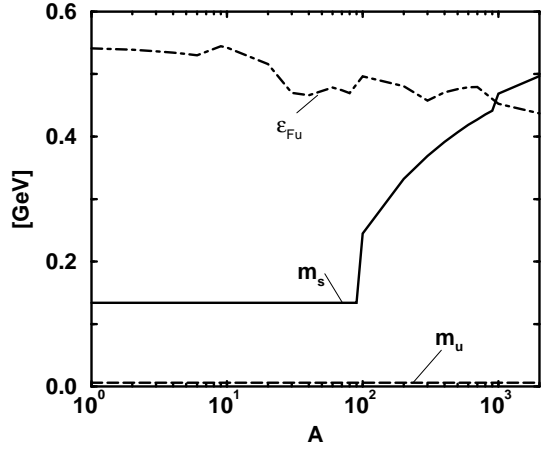}
\end{center}
\vspace*{-1.0cm} \caption{\small \baselineskip=0.5cm The $ud$ quark Fermi energy $\epsilon_{Fu}$ and the $s$ quark dynamical mass $m_{s}$ as functions of the baryon number $A$.}
 \label{fig : efu_ms_A}
\end{figure}

The chiral symmetry breaking of the $s$ quarks affects the stability of the strangelets.
The dynamical quark mass $m_{u}$ and $m_{s}$  of the $ud$ and $s$ quark are obtained as functions of the baryon number $A$ in Fig.~\ref{fig : efu_ms_A}.
The $ud$ quark Fermi energy $\varepsilon_{Fu}$ is shown in the same figure.
The chiral symmetry of the $ud$ and $s$ quarks is restored for $A \ltap 10^{3}$.
The $s$ quarks are contained in the ground state of the quark droplets, since the relation
\begin{eqnarray}
  m_{s} < \epsilon_{Fu},
  \label{eq : beta_condition}
\end{eqnarray}
is satisfied.
On the other hand, the chiral symmetry of the $s$ quarks is broken at $A \simeq 10^{3}$, where the dynamical $s$ quark mass is equal to the $ud$  quark Fermi energy.
Therefore, $s$ quarks do not appear in  the quark droplets with baryon number $A \gtap 10^{3}$.

Now, let us discuss the structure of the quark droplets for three ranges of the baryon number.

%%%%%%%%%%%%%%%%%%%%%%%%%%%%%%%%%%
%%%%%%%%%%%%%%%%%%%%%%%%%%%%%%%%%%

%%%%%%%%%%%%%%%%%%%%%%%%%%%%%%%%%%
%%%%%%%%%%%%%%%%%%%%%%%%%%%%%%%%%%

%%%%%%%%%%%%%%%%%%%%%%%%%%%%%%%%%%
%%%%%%%%%%%%%%%%%%%%%%%%%%%%%%%%%%

\subsubsection{The case of $ \bf 1 \le A \leq 4$}

% A=1 (baryon)

\begin{figure}[tbp]
\begin{minipage}{8cm}
\vspace*{0.0cm}
\mbox{ \small \bf (a)}
\centering
\includegraphics[width=8cm]{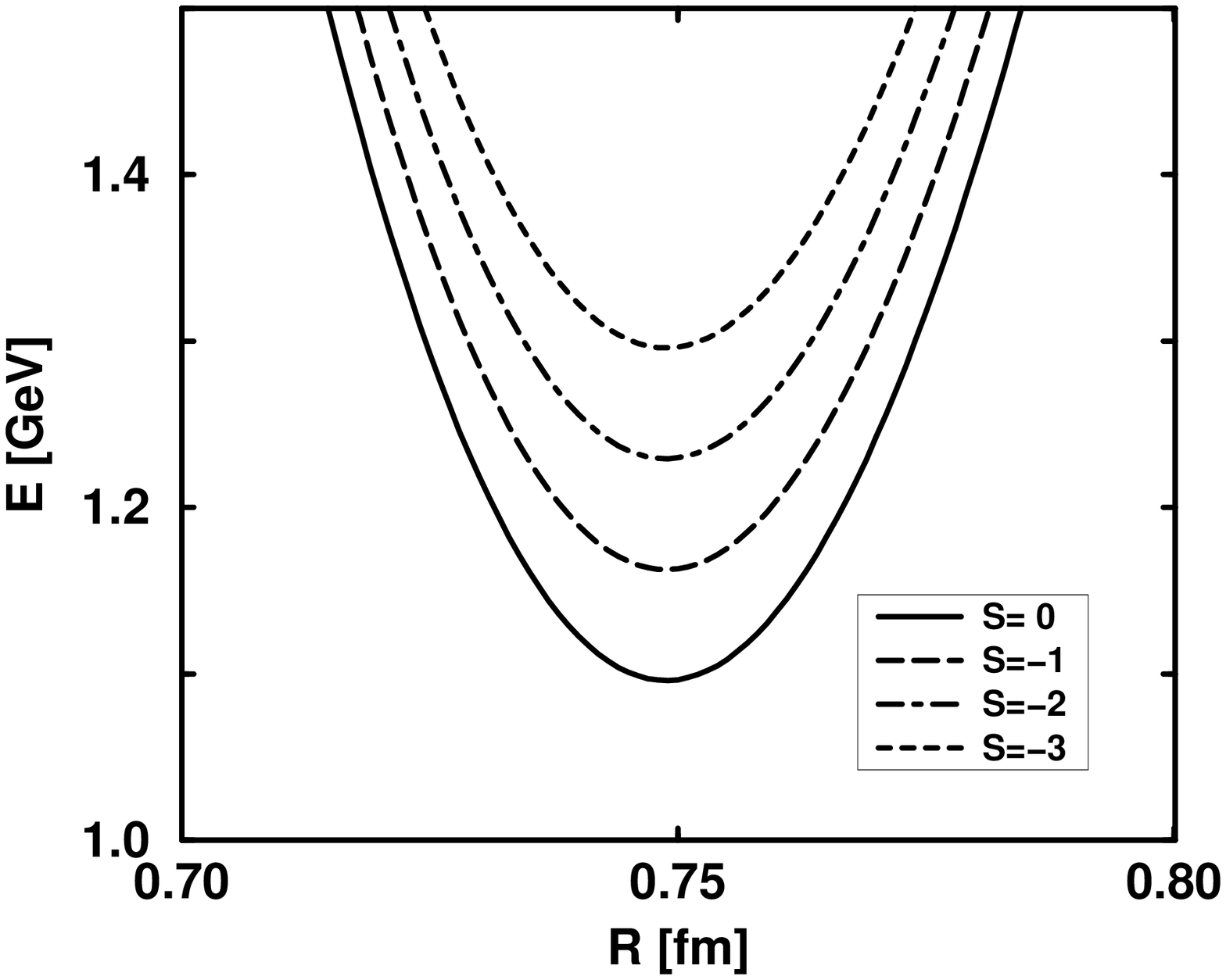}
\vspace{-0.0cm}
\end{minipage}
%%%%%%%%%%%%%%%%%%%%%%%%%%%%%%%%%
\begin{minipage}{8cm}
\mbox{ \small \bf (b)} \\
\centering
\includegraphics[width=5cm]{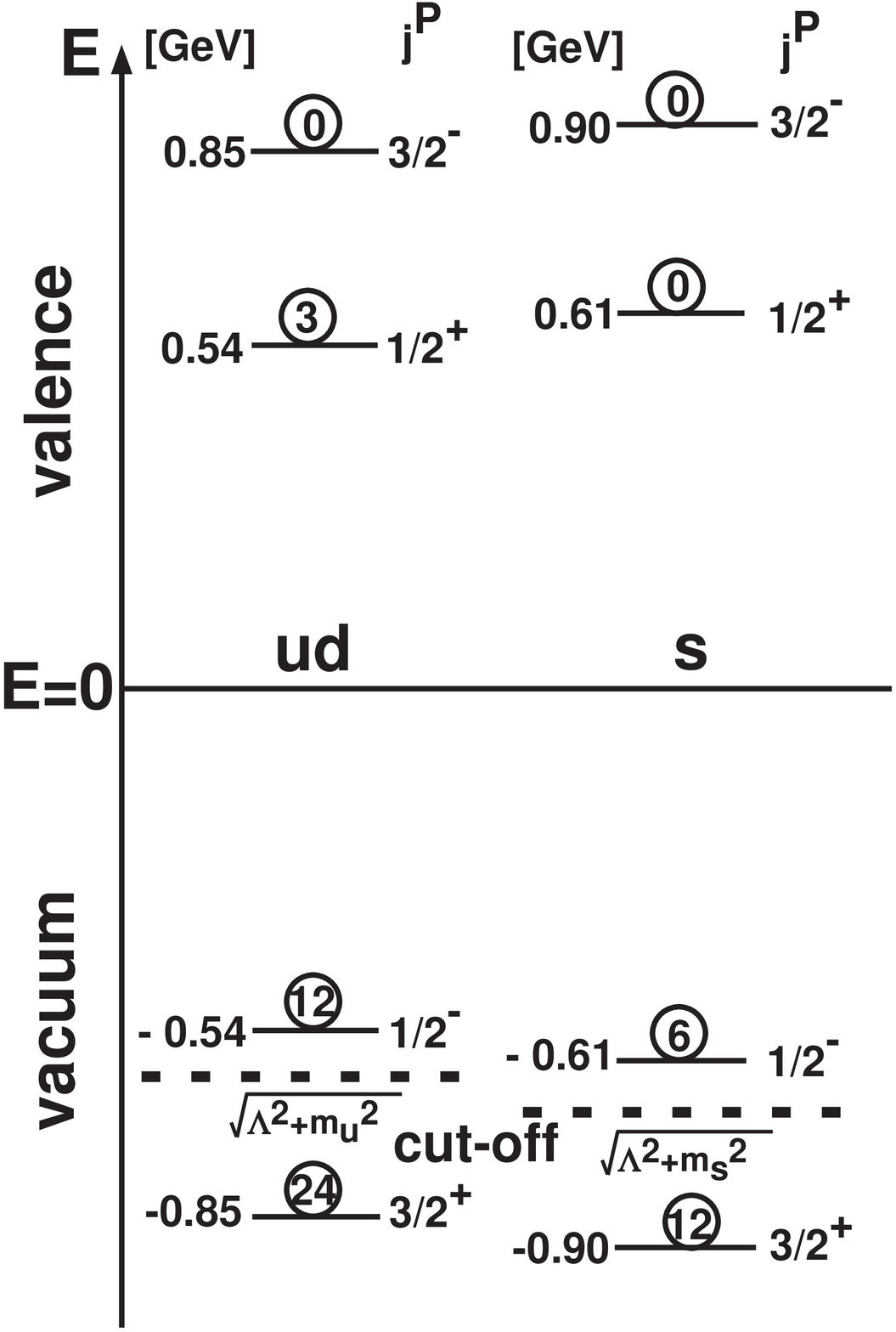}
\vspace{-0cm}
\end{minipage}
\caption{\small \baselineskip=0.5cm (a) The energies $E$ of a baryon ($A=1$)  for the strangeness $S=0, -1, -2$ and $-3$ as functions of the bag radius $R$. (b) The energy levels of eigenstate $J^{P}$ of the $u$, $d$ (left row) and $s$ (right row) quarks in the valence (upper) and the vacuum (lower) part in the bag of the baryon with strangeness $S=0$ and radius $R=0.75$ fm. The numbers in the circles are the number of the quarks occupying the levels.}
    \label{fig : figure_A1}
\end{figure}

We first consider a quark droplet with $A=1$, namely a baryon.
In Fig.~\ref{fig : figure_A1}(a), we show the energies $E$  for baryons with each strangeness $S\!=\!0$, $-1$, $-2$ and $-3$ as functions of the bag radius $R$.
There are local minima $E\!=\!1.10$, $1.16$, $1.23$ and $1.30$ GeV at bag radius $R=0.75$ fm for $S\!=\!0$, $-1$, $-2$ and $-3$, respectively.
The ground state is the nucleon with the mass $m_{N}=1.10$ GeV and the bag radius $R=0.75$ fm.
The inside of the baryon is the chirally restored phase  (see  Fig.~\ref{fig : efu_ms_A}).

%%%%%%%%%%%%%%%%%%%%%%%%%%%%%%%%%%%%%%%%%%%%%%%
%%%%%%%%%%%%%%%%%%%%%%%%%%%%%%%%%%%%%%%%%%%%%%%%%%

% vacuum structure in a bag
Now, we discuss the quark level structure in the baryon shown in Fig.~\ref{fig : figure_A1}(b), where we show the energies of the quark eigenstates in units of GeV on the left of the levels and the corresponding total angular momentum $j$ and parity $P$ on the right.
The numbers in the circles on the levels are the number of quarks that occupy the levels.
The dashed lines in the vacuum part indicate the momentum cut-off in the regularization.
The states below the cut-off are hidden in the shadow.

The three valence quarks occupy the state $1/2^{+}$ in the positive energy part of the $ud$ quark sector.
The state $1/2^{+}$ of the $s$ quark sector is not occupied, since its energy is larger than that of the $ud$ quark due to the difference of the current masses.
Therefore, the ground state of the baryon is non-strangeness state of $S=0$.
Concerning the vacuum quarks, only the $1/2^{-}$ state is included within the momentum cut-off as discussed in Eq.~(\ref{eq : function3}).
The next level of $3/2^{+}$ and more states existing below are not inside the cut-off $\Lambda$.

\begin{table}[tdp]
\begin{center}
\begin{tabular}{|c|c|c|c|}
\hline
     hadrons    &    $S$  & $E$ (cal.) [GeV] & $E$ (exp.) [GeV] \\
\hline
     $N$          &      0     &  1.10                 &    0.939                 \\ 
  $\Delta$    &      0     &   1.10                &    1.230                 \\
$\Lambda$ &     -1     &   1.16                &    1.115                \\
$\Sigma$    &     -1     &   1.16                &    1.190                 \\
$\Xi$           &     -2     &    1.23               &     1.314                 \\
$\Omega$  &     -3     &    1.30               &     1.672                  \\       
\hline
\end{tabular}
\end{center}
\caption{\small \baselineskip=0.5cm The energy of baryon with strangeness $S$. We denote $E$(cal.) and $E$(exp.) as calculated and experimental values, respectively.}
\label{tbl : baryon_experimental}
\end{table}%

%%%%%%%%%%%%%%%%
We stress that the shell structure in the vacuum part inside the bag is important for the baryon mass.
Indeed, the baryon mass was not obtained by the MRE method in \cite{Yasui}, because the MRE method was not appropriate to discuss the properties of a small size quark droplet due to the limitation of the approximation of this method.
%%%%%%%%%%%%%%%%
In the present method, despite the simplicity of the model, we have obtained the masses of $A=1$ baryons, especially of the nucleon reasonably well.

We have calculated masses of other SU(3) baryons.
The result is shown in Table \ref{tbl : baryon_experimental} as compared with experimental values.
We obtain $E_{\Lambda}=E_{\Sigma}=1.16$ GeV, $E_{\Xi}=1.23$ GeV and $E_{\Omega}=1.30$ GeV.
These masses are close to the experimental values.
In our calculation, the octet and decuplet baryons, for instance $N$ and $\Delta$, are degenerate; spin dependent residual interaction is responsible for the octet-decuplet splitting.
We check that the Gell-Mann--Okubo mass formula
\begin{eqnarray}
 2(N+\Xi) = 3 \Lambda + \Sigma
\end{eqnarray}
holds within $1.1$ \% in our numerical result.

Now, let us turn to the discussion for $A \ge 2$.
Since the $1/2^{+}$ state can accommodate twelve quarks at most, 
the ground state of the quark droplets with the baryon number $A=2$, $3$ and $4$ is still the non-strange state; the $ud$ quark droplets.
Then, we obtain the energy of the $ud$ quark droplets given by local minima as functions of the bag radius, as we discussed in the case $A=1$.
Let us see the resulting energy per baryon number $E/A$ of the quark droplets with baryon number $A$ in Fig.~\ref{fig : E_A}.
We obtain $E/A=1.36, 1.44$ and $1.48$ GeV for $A=2$, $3$ and $4$, respectively, which are larger than the nucleon mass $E=1.10$ GeV.
Therefore,  the $ud$ quark droplets of $2 \leq A \leq 4$ are not stable as compared with the $A$-nucleon system.
The reason for the instability of the $ud$ quark droplets is the increase of the effective bag constant $B$ for the bag radius $R$ for $0.7 \;\mbox{fm} \ltap R \ltap 1.05 \; \mbox{fm}$ as shown in Fig.~\ref{fig : B_R_A}(a).
For example, the quark droplets with baryon number $A=2$, $3$ and $4$  have the bag constant $B^{1/4}=0.077$, $0.10$ and $0.12$ GeV, respectively, while the baryon of $A=1$ has $B^{1/4}=0.036$ GeV.
Therefore, the energy per baryon number of the $ud$ quark droplets becomes larger than the baryon mass.

\subsubsection{The case of $ \bf 5 \le A \ltap 10^{3}$}

\begin{figure}[tbp]
\begin{minipage}{8cm}
\mbox{ \small \bf (a)}
\vspace*{0.0cm}
\centering
\includegraphics[width=8cm]{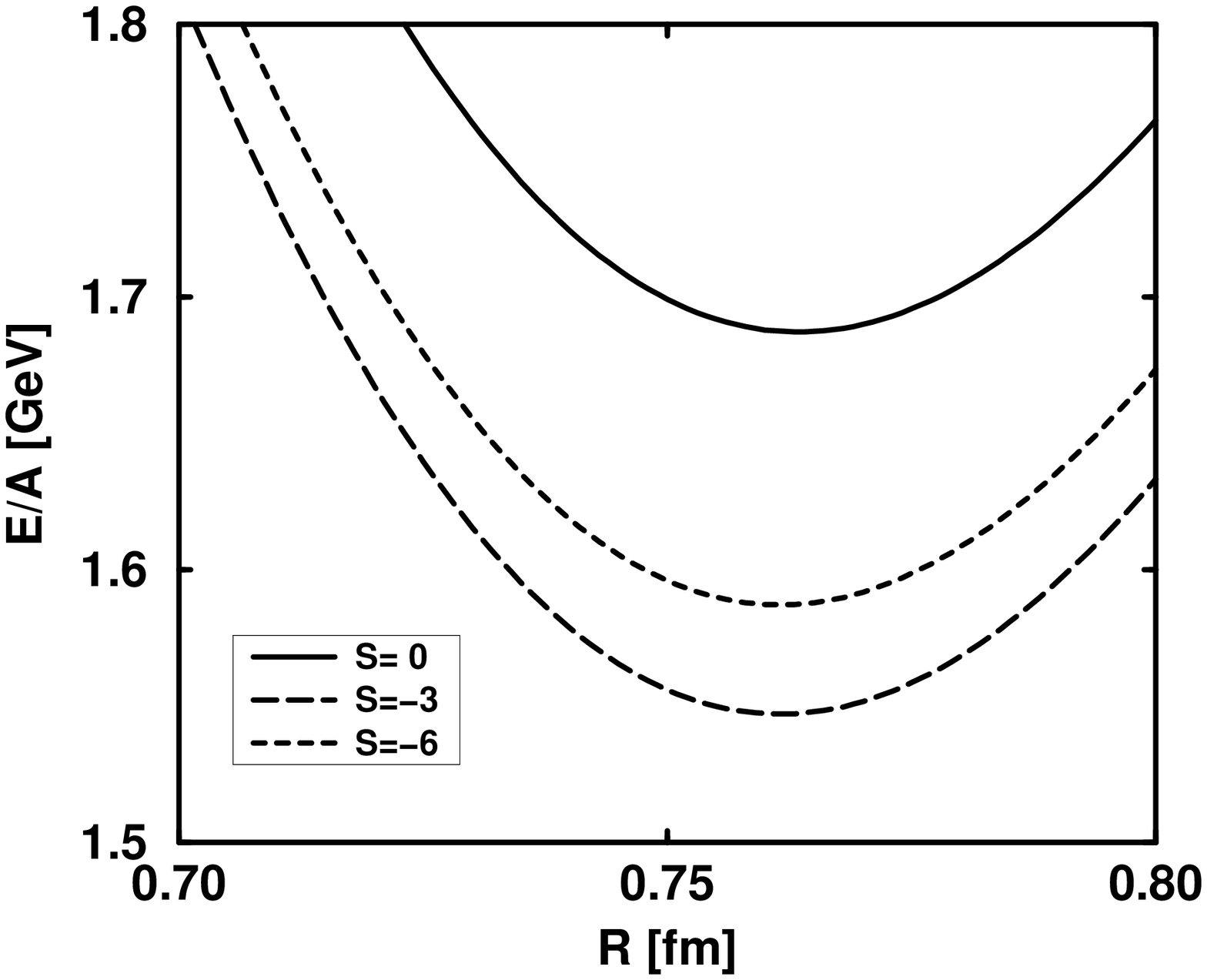}
\vspace{-0.0cm}
\end{minipage}
%%%%%%%%%%%%%%%%%%%%%%%%%%%%%%%%%
\begin{minipage}{8cm}
\mbox{ \small \bf (b)} \\
\centering
\includegraphics[width=5cm]{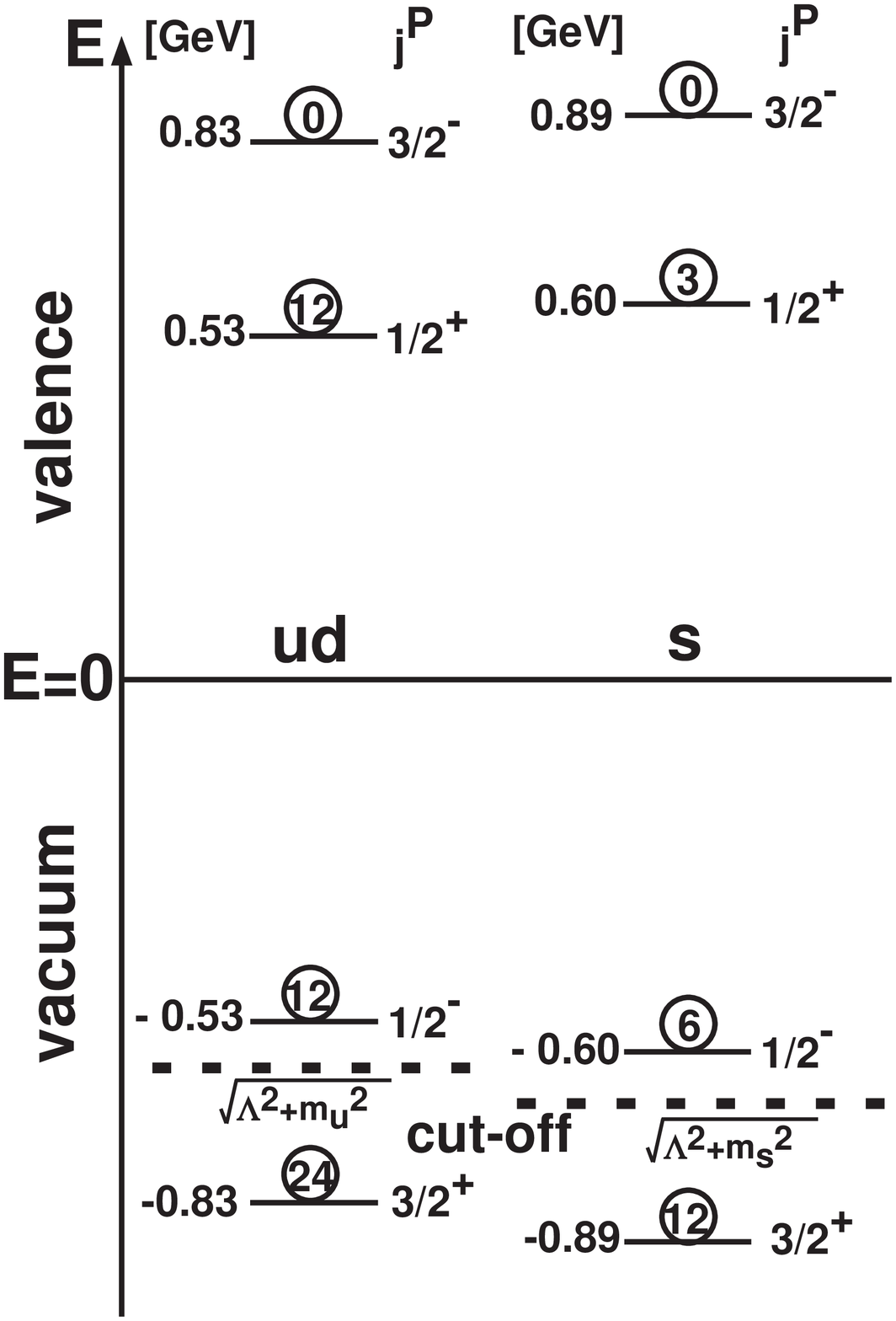}
\vspace{-0cm}
\end{minipage}
\caption{\small \baselineskip=0.5cm (a) The energies per baryon number $E/A$ of quark droplets with baryon number $A=5$  for the strangeness $S=0, -3$ and $-6$ as functions of the bag radius $R$. (b) The energy level of eigenstate $J^{P}$ of the $u$, $d$ (left row) and $s$ (right row) quarks in the valence (upper) and the vacuum (lower) part in the bag of baryon number $A=5$, strangeness $S=-3$ and radius $R=0.76$ fm.  The number in the circle is the number of the occupying quarks in each state.}
    \label{fig : figure_A5}
\end{figure}

% A=5
Next, let us discuss the quark droplets with baryon number $A \geq 5$.
For these baryon numbers, we obtain the strangelets as the ground state of the quark droplets.
Let us investigate the case of $A=5$.
In Fig.~\ref{fig : figure_A5}(a), we plot the energies per baryon number $E/A$ of the quark droplets of $A=5$ for strangeness $S=0$, $-3$ and $-6$ as functions of the radius $R$.
At the local minima, we obtain $E/A = 1.68, 1.54$ and $1.58$ GeV for strangeness $S = 0, -3$ and $-6$, respectively.
The most stable state is the strangelet with the finite strangeness $S = -3$.
The bag radius is $R=0.76$ fm.
We can understand this result of the finite strangeness by considering the structure of the valence quark levels in the bag as shown in Fig.~\ref{fig : figure_A5}(b).
Here, we use the current quark masses to obtain the energy levels, since chiral symmetry is still restored at $A=5$ (see  Fig.~\ref{fig : efu_ms_A}).
The twelve quarks occupy fully the lowest state $1/2^{+}$ of the $ud$ quark sector.
We may put the remaining three quarks in the next level $3/2^{-}$ in $ud$ quark sector.
However, the energy $E_{3/2^{-}}=0.83$ GeV of the $ud$ quark sector is larger than the energy $E_{1/2^{+}}=0.60$ GeV in the $s$ quark sector.
Therefore,  the state $1/2^{+}$ of the $s$ quark is occupied by the remaining three quarks.
Therefore, the ground state of the quark droplet with baryon number $A=5$ is strangelets with strangeness $S=-3$.

We have investigated the energy of the strangelet for larger $A$, and found that the strangelets can be the ground state as quark droplets with $A \ltap 10^{3}$.
From Fig.~\ref{fig : E_A}, the energies per baryon number $E/A$ of the strangelets of $5 \le A \ltap 10^{3}$ are $1.5$ GeV $ \ltap E/A \ltap$ 1.8 GeV, which are larger than the nucleons mass $E_{N}=1.10$ GeV and the Lambda mass $E_{\Lambda}=1.16$ GeV.
Therefore, the strangelets are not stable against normal nuclei and hypernuclei.

%%%%%%%%%%%%%%%%%%%%%%%%%%%%%%%%%%%%%%%%%%%%%%%
%%%%%%%%%%%%%%%%%%%%%%%%%%%%%%%%%%%%%%%%%%%%%%%%%%

\subsubsection{The case of $ \bf A \gtap 10^{3}$}

For the baryon number $A \gtap 10^{3}$, the ground state of the quark droplet is not strangelet, but the $ud$ quark droplet.
This is because the chiral symmetry in the $s$ quark sector is broken.
As shown in Fig.~\ref{fig : efu_ms_A}, the dynamical mass $m_{s}$ of the $s$ quark becomes larger than the $ud$ quark Fermi energy $\epsilon_{Fu}$ for $A\gtap 10^{3}$.
There, the condition for the weak decay (\ref{eq : beta_condition}) is not fulfilled.
Therefore, the $s$ quarks cannot appear in the ground state of the quark droplets.

%\vspace{1cm}

From the discussion above, it is shown that the baryon number of stable strangelets is limited in $5 \le A \ltap 10^{3}$.
This conclusion is qualitatively different from the result in the MIT bag model;  all strangelets of the baryon number, say $A \gtap 10$, are the most stable state as compared with nuclei \cite{Farhi, Madsen}.
In contrast, our result by the NJL+MIT bag model indicates that the strangelet can exist as a quasi-stable state rather than the true ground state.

\subsubsection{Choice of the parameter $a$}

Up to now, we have used a parameter set reproducing the properties of the infinite volume vacuum.
However, the values of the diffuseness parameter $a$ in the Lorentzian regularization (\ref{eq : regularization}) 
cannot be determined only from the properties of the infinite volume system, because such a system is insensitive to the parameter $a$.
In order to determine $a$, let us look at the $a$-dependence of the nucleon mass and radius.

We show nucleon mass $E$, the bag radius $R$ and the effective bag constant $B$ for several diffuseness parameter $a$ in Table~\ref{tbl : tbl1}.
As the diffuseness parameter changes from $a=20$ (smooth) to $25$ (sharp), the baryon mass $E$ and the effective bag constant $B^{1/4}$ changes from $E=1.32$ GeV to $0.83$ GeV and from $B^{1/4}=0.177$ GeV to $-0.186$ GeV.
For $a=22$, we find $E=1.10$ GeV which is the average value of the masses of the nucleon and the delta.
This is the reason why we have used $a=22$ in the preceding discussions.
On the other hand, the nucleon radius remains around $R \simeq 0.75$ fm for the range of $a$ shown in Table~\ref{tbl : tbl1}.
We also checked that the Gaussian regularization with an appropriate parameter set.
However, the nucleon mass was not reproduced in the Gaussian type.

The energy of the quark droplet in the present framework depends on the parameter $a$ particularly for small $A$ region.
However, the present result for $A \gtap 10$ is rather parameter independent, and the result of $A=1$ should be like the one of the nucleons and the delta, as it was reproduced in the present treatment by suitably choosing $a$.
Therefore, we expect a smooth interpolation of the two points, for instance ($A=10$, $E/A=1.8$ \mbox{GeV}) and ($A=1$, $E=1.10$ \mbox{GeV}) of Fig.~\ref{fig : E_A}, although we need further refined treatment for the detailed behavior for small $A$ region.

\begin{table}[tdp]
\begin{center}
\begin{tabular}{|c|c|c|c|c|c|c|}
\hline
 & \multicolumn{3}{ c | }{nucleon} & \multicolumn{3}{ c | }{$ud$ quark matter} \\
\cline{2-7}
    \multicolumn{1}{| c |}{\raisebox{1.5ex}[0pt]{$a$}}    & $E$ [GeV] & $R$ [fm] & $B^{1/4}$ [GeV] & $\varepsilon/n_{B}$ [GeV] & $n_{B}/n_{B0}$ & $B^{1/4}$ [GeV] \\
\hline
   20   &     1.32      &  0.747  &    0.177    & 1.121 & 2.6 & 0.174  \\ 
   21   &     1.20      &  0.748  &    0.147    & 1.117 & 2.6 & 0.174  \\
   22   &     1.10      &  0.749  &    0.0362  & 1.114 & 2.6 & 0.173  \\
   23   &     1.00      &  0.749  &  -0.144    & 1.112 & 2.5 & 0.173  \\
   24   &     0.90      &  0.749  &  -0.167    & 1.110 & 2.5 & 0.173  \\
   25   &     0.83      &  0.750  &  -0.186    & 1.107 & 2.5 & 0.173  \\
\hline
\end{tabular}
\end{center}
\caption{\small \baselineskip=0.5cm We show nucleon mass $E$, the radius $R$ and the effective bag constant $B$ of a single baryon for several diffuseness parameter $a$ in the left side. In the right side, the energy per baryon, the stable density and the effective bag constant of $ud$ quark matter are shown.}
\label{tbl : tbl1}
\end{table}%

\subsubsection{Other quantities}

\begin{figure}[ptb]
\begin{center}
\includegraphics[width=8cm, height=8cm, keepaspectratio, clip]{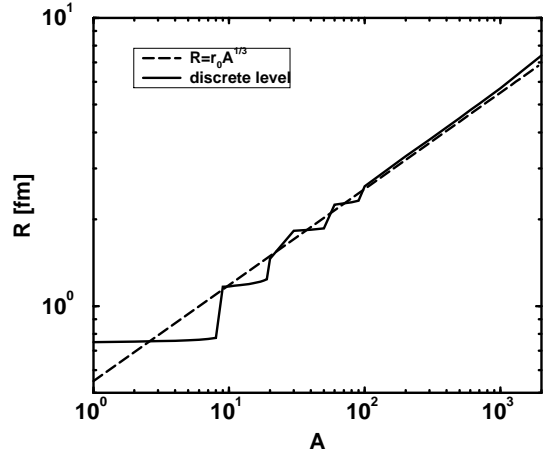}
\end{center}
\vspace*{-1.0cm} \caption{\small \baselineskip=0.5cm The solid line is the radius $R$ of strangelets as a function of baryon number $A$. The dashed line represents $R=r_{0}A^{1/3}$ with $r_{0}=0.56$ fm. }
 \label{fig : R_A}
\end{figure}

% radius of strangelet
We discuss radii and baryon densities of strangelets.
In Fig.~\ref{fig : R_A}, we show the bag radius $R$ of the strangelets as a function of the baryon number $A$.
The solid line shows the bag radius which is obtained by the discrete level calculation.
We fit our numerical result by an approximate relation $R=r_{0}A^{1/3}$ with a constant $r_{0}$, in the dashed line.
Here, we obtain the constant $r_{0}=0.56$ fm by fitting our result. 
Then, the averaged baryon number density is $n_{B}=A/V \simeq 8 n_{B0}$, where $n_{B0}=0.17$ $\mbox{fm}^{-3}$ is the normal nuclear matter density.
Hence, the strangelets are more compact objects as compared with ordinary nuclei.

A strangelet existing as a locally stable state has a small electric charge as compared with the normal nuclei.
The electric charge is given by $Q = (2/3) N_{u} - (1/3) N_{d} - (1/3) N_{s} = 0.5(1-3r_{s})A$.
From Fig.~\ref{fig : rs_A}, we obtain the charges per baryon number $0 \le Q/A \ltap 0.3$ for the strangelets of $5 \leq A \ltap 10^{3}$, which are smaller than the values in the normal nuclei $Q/A \sim 0.5$.
Therefore, the strangelets would be observed as heavy particles with small electric charges.

%%%%%%%%%%%%%%%%%%%%%%%%%%%%%%%%%%
%%%%%%%%%%%%%%%%%%%%%%%%%%%%%%%%%%

\subsection{Bulk quark matter}

In this subsection, we discuss the stability of the bulk quark matter in comparison with the nuclear matter.
In \cite{Buballa96, Buballa98, Buballa99}, it was shown by using the NJL model that the ground state of the quark matter is not the strange matter, but the $ud$ quark matter.
This result agrees with our present result that the quark droplets with large baryon number $A \gtap 10^{3}$ are the $ud$ quark droplets, not the strangelets.
In \cite{Buballa96, Buballa98, Buballa99}, however, it was also not possible to compare the stability of the $ud$ quark matter with that in the nuclear matter, since the baryon mass was not derived in the NJL model in their treatment.
In contrast, our present model can reproduce the baryon mass, which can be compared with the energy of the bulk quark matter.

We consider that the bulk quark matter in the present model (\ref{eq : NJL_bag}).
This is obtained either  by neglecting confinement term in Eq.~(\ref{eq : NJL_bag}), or by taking the limit of the infinite radius.
Then, in the mean field approximation (\ref{eq : constituent_mass}) and (\ref{eq : dynamical_mass}), the energy density of the bulk quark matter is given by \cite{Buballa96, Buballa98, Buballa99}
\begin{eqnarray}
\varepsilon = \sum_{q=u,d,s} 
                   \left[
               				   \frac{\phi_{q}^{2}}{4G} 
						   + \nu \int^{p_{Fq}} \frac{d^{3}p}{(2\pi)^{3}} g(p/\Lambda) \sqrt{p^{2}+(m_{0q}+\phi_{q})^{2}} \right] - \varepsilon_{0}.
\label{eq : NJL_bulk_energydensity}
\end{eqnarray}
The quark number density is
\begin{eqnarray}
n_{q} = \nu \int_{0}^{p_{Fq}} \frac{p^{2}dp}{2\pi^{2}},
\end{eqnarray}
for each flavor $q\!=\!u$, $d$ and $s$.
In Eq.~(\ref{eq : NJL_bulk_energydensity}), instead of the sharp cut-off, we have a smooth cut-off by $g(p/\Lambda)$.

\begin{figure}[tbp]
\begin{minipage}{8cm}
\vspace*{0.0cm}
\mbox{ \small \bf (a)}
\centering
\includegraphics[width=8cm]{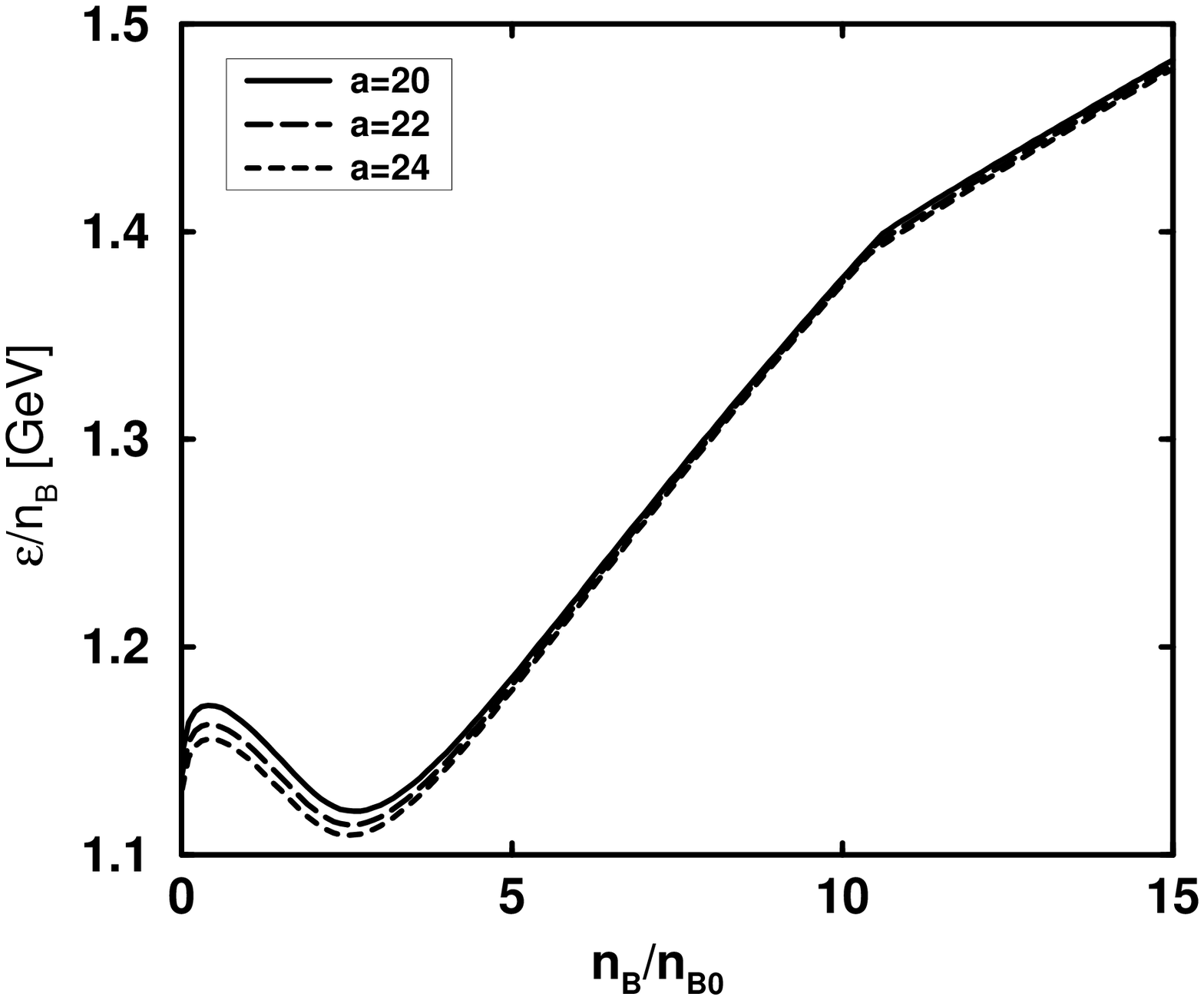}
\vspace{-0.0cm}
\end{minipage}
%%%%%%%%%%%%%%%%%%%%%%%%%%%%%%%%%
\begin{minipage}{8cm}
\mbox{ \small \bf (b)} \\
\centering
\includegraphics[width=8cm]{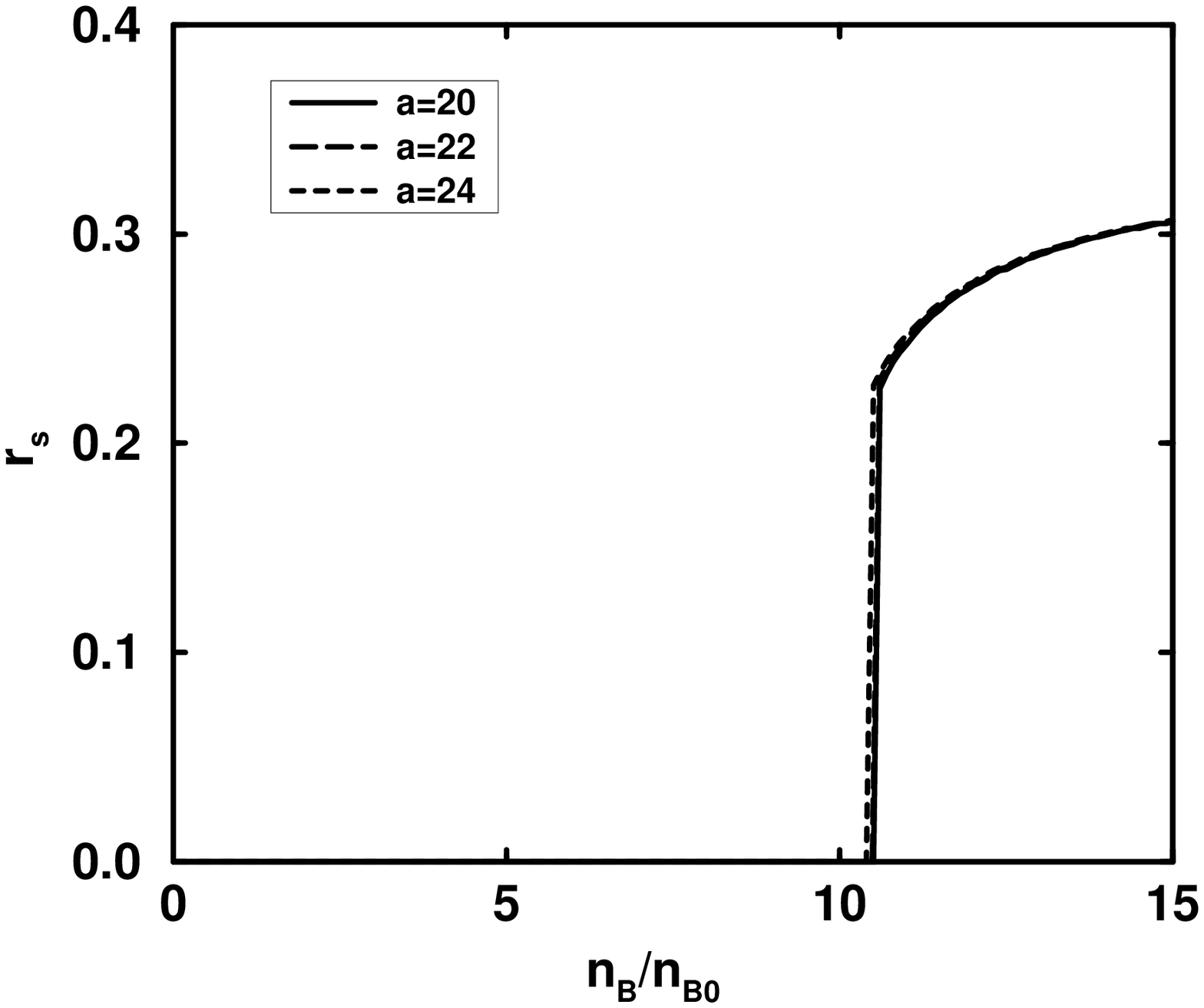}
\vspace{-0cm}
\end{minipage}
\caption{\small \baselineskip=0.5cm (a) The energy per baryon number $\varepsilon/n_{B}$ as functions of the baryon number density $n_{B}$. (b) The strangeness fraction $r_{s}$ as functions of the baryon number density $n_{B}$. $n_{B0}=0.17$ $\mbox{fm}^{-3}$ is normal nuclear matter density. $a$ is diffuseness parameter in the momentum regularization.}
    \label{fig : E_dq_a}
\end{figure}

In Fig.~\ref{fig : E_dq_a}(a), we show the energy per baryon number $\varepsilon/n_{B}$ as functions of the baryon number density $n_{B}\!=\!(n_{u}\!+\!n_{d}\!+\!n_{s})/3$ for the diffuseness parameters $a=20$, $22$ and $24$.
We also show the strangeness fraction $r_{s}$ obtained by a variation of the energy per baryon $\varepsilon/n_{B}$ in Fig.~\ref{fig : E_dq_a}(b).
The ground state of the quark matter is $ud$ quark matter for lower density $n_{B} \ltap 10.5 n_{B0}$, because the chiral symmetry of $s$ quark sector is broken at the vacuum and the lower density.
On the other hand, the ground state of the quark matter is strange matter for higher density $n_{B} \gtap 10.5 n_{B0}$, since the chiral symmetry of $s$ quark sector is restored.
In fact, the strangeness fraction takes a finite value at the density $n_{B} \gtap 10.5 n_{B0}$.

From Fig.~\ref{fig : E_dq_a}(a), there are stable states of the quark matter around around $n_{B}/n_{B0}=2.5$.
This is the $ud$ quark matter, not the strange matter.
There, we obtain the energy per baryon number and the baryon number density of the stable $ud$ quark matter.
We summarize our results in Table \ref{tbl : tbl1}.
The energy per baryon is $\varepsilon/n_{B}=1.10-1.12$ GeV, the stable density is $n_{B}/n_{B0}=2.5-2.6$ , and the effective bag constant is $B^{1/4}=0.173-0.174$ GeV.
These quantities in the bulk quark matter are not affected by the diffuseness parameter $a$ so much as in the baryon.

Let us compare the energy $E$ of baryon and the energy per baryon number $\varepsilon/n_{B}$ in the $ud$ quark matter.
For the diffuseness parameter $a \ge 22$, the $ud$ quark matter is unstable against the baryon.
On the other hand, for the diffuseness parameter $a \le 21$, the baryon mass $E$ exceeds the energy per baryon number $\varepsilon/n_{B}$ of the $ud$ quark matter.
Therefore, for the diffuseness parameter $a \le 21$, we can expect that the quark matter can be stable state rather than the nuclear matter.
From Fig.~\ref{fig : E_A}, the $ud$ quark droplets in the limit of the large baryon number can be more stable than the baryon system, when the energy per baryon of the $ud$ quark matter is smaller than the baryon mass.
Even in that case, however, the strangelets are not absolutely stable objects due to the large energy per baryon as compared with the baryon mass.

%%%%%%%%%%%%%%%%%%%%%%%%%%%%%%%%%%%%%%%%%%%%%%%
%%%%%%%%%%%%%%%%%%%%%%%%%%%%%%%%%%%%%%%%%%%%%%%%%%

% decay mode of strangelets

\section{Conclusion}

In this paper, we discussed the stability of the quark droplet by considering dynamical chiral symmetry breaking in a finite volume region of the droplet.
We considered chiral symmetry breaking by using the NJL model supplemented by the MIT bag boundary condition for quark confinement.
The boundary condition prepared the basis set in the bag.
Then, we obtained the energy per baryon number of the quark droplets by solving self-consistently the equation of state in the mean field approximation.
Masses and radii of various baryons were also discussed in this model.

We obtained the result that the strangelets are stable as quark droplets for the baryon number $5 \leq A \ltap 10^{3}$.
This is because the chiral symmetry of the $s$ quark sector is restored for these baryon numbers.
We also obtained the baryon mass $E_{N}=1.10$ GeV and the radius $R=0.75$ fm as a quark droplet with baryon number $A=1$, which was identified with the nucleon.
However, the energy per baryon number of strangelets are larger than the baryon mass.
Therefore, we conclude that the strangelets can not be the ground state in the present framework.
We also discussed the stability of the bulk quark matter against the nuclear matter.
The baryon mass depends on the diffuseness parameter in the Lorentzian regulator, while the energy of the bulk matter did not.
By comparing baryon mass and the energy per baryon number in the stable $ud$ quark matter, it was concluded that the $ud$ quark matter could be more stable than the baryon.
 
We also confirmed that the MRE method reproduces the results of the present calculation using discrete levels directly for the baryon number $A \gtap 10$.
This justifies the use of the MRE for larger $A$, while for smaller $A$ the shell structure of the discrete levels becomes important.

We have still some possibilities that the energy of strangelets is modified by other effects which were not fully considered in our discussion.
As one of the most important terms, we need to consider the 't Hooft term for explicit $U(1)_{A}$ breaking.
We may need to consider the meson clouds around the bag also.
In fact, from the view of chiral symmetry, the $\pi$ and $K$ mesons should be introduced at the bag surface in order to recover chiral symmetry broken at the bag surface.
In the present work, we neglected them.
This approximation is expected to be valid for systems with large baryon number.
For instance, in the previous studies of the chiral bag model, it was shown that the pion effect for $A=2$ is significantly smaller than the nucleon ($A=1$) \cite{Vento_Rho, Takashita}.
Therefore, at least for $A \gtap 2$, we expect that the present result will not be affected so much by the meson cloud.
Nevertheless, the inclusion of the meson cloud in the model is very important in particular for the interpolation of the present analysis to the small $A$ region ($1 \le A \ltap 10$).
We are currently working out such a problem. 
Another interesting effect is the diquark correlation in the quark droplets \cite{Madsen01, Amore, Kiriyama}.
In recent studies of high density matter, it has been discussed that the true ground state of the quark matter may be color superconductivity state.
Such effects would be necessary to further development of the study of the strangelets.

%%%%%%%%%%%%%%%% Reference %%%%%%%%%%%%%%

\end{document}